\documentclass{emulateapj}

\usepackage{amsmath}
\usepackage{epstopdf}

\newcommand{\be}{\begin{equation}}
\newcommand{\ee}{\end{equation}}

\shorttitle{GRB Temporal Variability}
\shortauthors{Golkhou \& Butler}

\slugcomment{Accepted to ApJ}

\begin{document}

\setlength{\pdfpageheight}{\paperheight}
\setlength{\pdfpagewidth}{\paperwidth}

\title{Uncovering the Intrinsic Variability of Gamma-ray Bursts}

\author{V. Zach Golkhou\altaffilmark{1,2}
\& Nathaniel R. Butler\altaffilmark{1,2}}

\altaffiltext{1}{School of Earth and Space Exploration,
Arizona State University, Tempe, AZ 85287, USA}
\altaffiltext{2}{Cosmology Initiative,
Arizona State University, Tempe, AZ 85287, USA}

\begin{abstract}
We develop a robust technique to determine the minimum variability timescale for Gamma-ray Burst light curves, utilizing Haar wavelets.  Our approach averages over the data for a given GRB, providing an aggregate measure of signal variation while also retaining sensitivity to narrow pulses within complicated time-series.
In contrast to previous studies using wavelets, which simply define the minimum timescale in reference to the measurement noise floor, our approach identifies the signature of temporally-smooth features in the wavelet scaleogram and then additionally identifies a break in the scaleogram on longer timescales as signature of a true, temporally-unsmooth light curve feature or features. We apply our technique to the large sample of \textit{Swift} GRB Gamma-ray light curves and for the first time -- due to the presence of a large number of GRBs with measured redshift -- determine the distribution of minimum variability timescales in the source frame. We find a median minimum timescale for long-duration GRBs in the source frame of $\Delta t_{\rm min}=0.5$ s, with the shortest timescale found being on the order of 10 ms.   This short timescale suggests a compact central engine ($3 \times 10^3$ km).  We discuss further implications for the GRB fireball model and present a tantalizing correlation between minimum timescale and redshift, which may in part be due to cosmological time-dilation.
\end{abstract}


\maketitle

\section{Introduction}
\label{sec:intro}

Gamma-Ray Burst (GRB) light curves show a remarkable morphological diversity. While a significant number of bright long bursts ($\sim15\%$) exhibit a single smooth pulse structure, in most cases GRBs appear to be the result of a complex, seemingly random distribution of several pulses. Burst pulses are commonly described as having fast-rise exponential-decay (FRED) shapes \citep[e.g.,][]{1996ApJ...473..998F}. Parameterized analyses of pulse profiles have shown broad log-normal distributions among different bursts and even within a single burst \citep[see, e.g.,][]{1996ApJ...459..393N,2012ApJ...744..141B}.

Several approaches have been utilized to characterize the distribution of power versus timescale for GRBs and other astrophysical sources.  These include structure function (SF) analyses \citep{1994ApJ...433..494T,1994MNRAS.268..305H,1996A&A...306..395C,1997MNRAS.286..271A}, autocorrelation function (ACF) analyses \citep{1993ApJ...408L..81L,1995ApJ...448L.101F,1996ApJ...464..622I,2004A&A...418..487B,2012ApJ...749..191C}, and Fourier power spectral density (PSD) analyses \citep{2000ApJ...535..158B,2001ApJ...557L..85C,2010ApJ...722..520A,2012MNRAS.422.1785G,2013MNRAS.431.3608D}.
In principle, the ACF contains the same information as the PSD, since one is the Fourier transform of the other \citep[the Wiener-Khinchin theorem,][]{chatfield2003analysis}.  The SF is mathematically very similar to the ACF.

As \citet{2002MNRAS.329...76H} summarizes, the SF, ACF, and PSD -- when calculated for a given dataset -- are not completely equivalent because of time-windowing effects and the presence of measurement noise.
For long runs of evenly spaced data, the PSD is used in preference to the ACF, as it can be easier to interpret and understand errors. In cases of short or inhomogeneous data sets, the ACF can provide a more stable measurement. However, as ACF values at different time lags are not statistically independent of each other, the ACF interpretation may not be simple.

\begin{figure}
\begin{center}
\includegraphics[width=.3\textwidth]{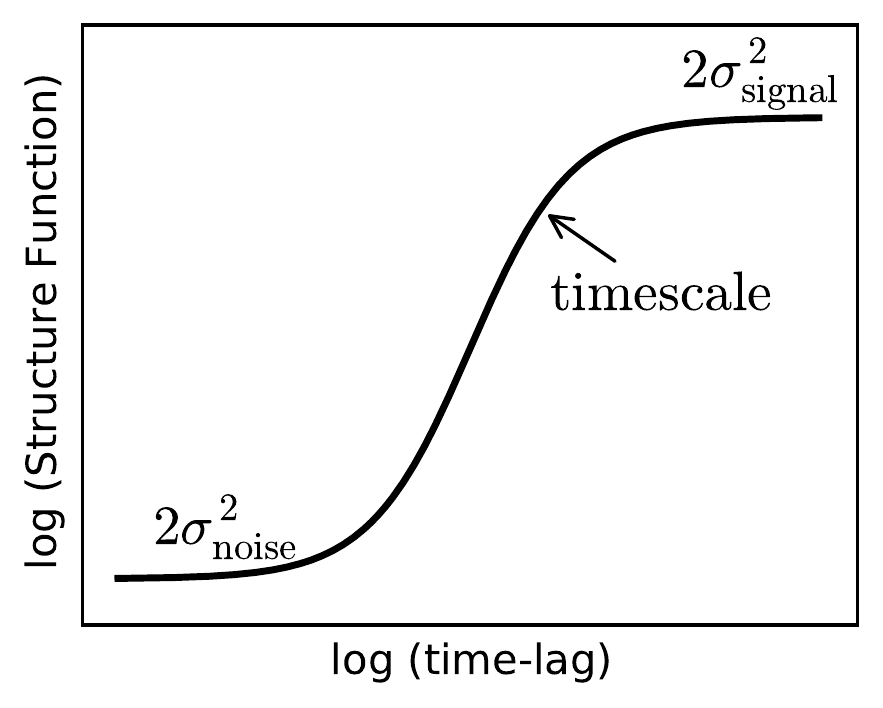}
\caption{\small 
Schematic showing a typical SF for a time-series, from \citet{1992ApJ...396..469H}. At short lag-times, the SF flattens due to the measurements error. At long lag-times, the SF again flattens out at a level corresponding to the total variance in the signal.  Between these lag-times, the slope of the SF depends on the noise properties of the signal and can be used to identify timescales of interest.}
\label{fig:sfp}
\end{center}
\end{figure}

The first-order SF was introduced in astronomy by \citet{1985ApJ...296...46S}.
It has been widely used in the analysis of quasar light curves \citep[e.g.,][]{1994ApJ...433..494T,2002MNRAS.329...76H} and microlensing statistics \citep[e.g.,][]{2001MNRAS.320...21W}.
Compared to power-spectral analyses, the SF approach is less dependent on the time sampling \citep{1999ASPC..159..293P}.  
Following these studies, we define the first-order SF as a measure of the mean square difference of a signal $X(t)$ on timescale (or ``lag'') $\tau$: 
\be
{\rm SF}(\tau)=\left\langle [X(t)-X(t+\tau)]^2 \right\rangle
\label{Eq:sf}
\ee
Here, $\left\langle . \right\rangle$ denotes an averaging over $t$.
In Figure \ref{fig:sfp}, we reproduce the typical shape of an SF, from \citet{1992ApJ...396..469H}.

We will be primarily interested below in using the SF to infer the shortest timescale at which a GRB exhibits {\it uncorrelated temporal variability}.  In a seminal study, \citet{2000ApJ...537..264W} \citep[and more recently,][]{2013MNRAS.432..857M} utilize Haar wavelet scaleograms to measure minimum timescales. 
Wavelets are a set of mathematical functions, which form an orthonormal basis to compactly describe narrow time features \citep[e.g.,][]{1992tlw..conf.....D,1994ApJ...424..540N,1997ApJ...483..340K,2013ApJ...764..167S}. Making the connection between the Haar wavelet scaleogram and the SF, as we do below in mathematical detail, sheds new light on prior work, allowing for a more rigorous analysis and better physical interpretation of the signal power versus timescale.  We also exploit the large sample of \textit{Swift} GRBs with measured redshifts to perform this analysis, for the first time, in the GRB source frame.

A general feature we observe in our scaleograms, provided there is sufficient signal-to-noise ratio (SNR), is a linear rise phase relative to the Poisson noise floor on the shortest timescales (see, e.g., Figures \ref{fig:pick_burst} and \ref{fig:simburst_30s}).  We take this to indicate a typical smoothness on the shortest observed timescales.
We thus make an essential distinction -- not made in prior studies -- between correlated variability (i.e., smooth or continuous) and uncorrelated variability (e.g., pulses or changes in sign).  For example, an exponentially decaying GRB light curve pulse with a fairly long time constant (say 100 s) will still exhibit power (i.e., yield a non-zero SF) on much shorter timescales (say 1 s), provided the SNR is sufficiently large for this to be measured.  In contrast, the meaningful timescale (in this case $\approx$ 100 s and not 1 s) is the shortest timescale at which the signal becomes uncorrelated.

A simple Taylor expansion of the SF assuming a temporally-smooth signal $X(t)$, shown in Equation \ref{Eq:taylor_expn}, elucidates how the minimum timescale for uncorrelated variability is connected to the scaleogram linear-rise phase.
\be
X(t+\tau) = X(t) + \tau X'(t)|_{\tau} + ...,
\label{Eq:taylor_expn}
\ee
Substituting Equation \ref{Eq:taylor_expn} into Equation \ref{Eq:sf} and ignoring higher order terms produces Equation \ref{Eq:prop}:
\be
\sqrt{{\rm SF}(\tau)} \propto \tau
\label{Eq:prop}
\ee
which shows that for timescales where the signal is smoothly varying, we expect a linear dependence on the time lag $\tau$.
When the variation becomes non-smooth, SF flattens, providing a signature of the true GRB minimum timescale.  Previous studies \citep{2000ApJ...537..264W,2013MNRAS.432..857M} -- which overlook the importance of the ${\rm SF} \propto \tau$ region -- incorrectly interpret the GRB minimum as the shortest timescale at which the SF is first non-zero (after subtracting the measurement noise level), potentially under-estimating the true variability timescales.

In this paper, we begin with a more detailed description of our method -- the Haar wavelet structure function -- which exploits a non-decimated, discrete Haar wavelet transform to estimate the SF and, in turn, the minimum variability timescale for a large number GRBs.  We discuss the robustness of the structure function in extracting this timescale even in the case of complex GRBs containing multiple, overlapping pulses and we demonstrate self-consistency as the SNR is varied.  Next, we apply the methodology to the full sample of GRBs observed by \textit{Swift} BAT, summarizing the derived timescales for the population in the observer and GRB source frames.  We conclude by discussing how these minimum variability timescales can elucidate the GRB central engine, help constrain models for the emission mechanism, and potentially also enable a measurement of cosmological time-dilation.

\begin{figure}
\begin{center}
\includegraphics[width=.49\textwidth]{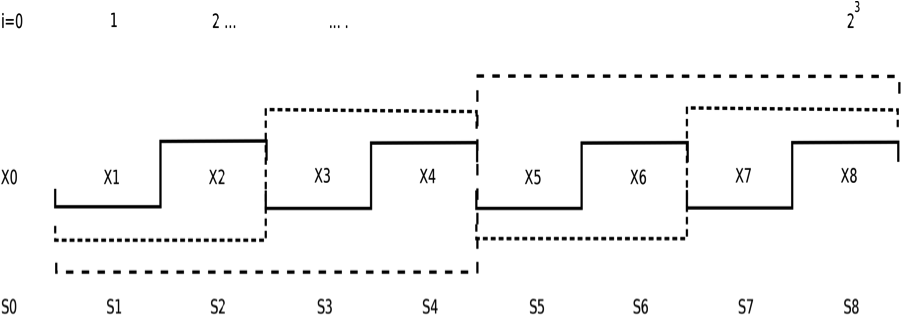}
\caption{\small 
A schematic representation of relation between the Haar wavelet coefficients (Equation \ref{Eq:coef}) and the first-order structure function. The Haar mother wavelet as shown step function (with different style) operates using scaling and dilation on a time-series: $\{X_i | i=0..8\}$. $S_i = \sum_{0}^{i}{X_i}$.}
\label{fig:ep_pred2}
\end{center}
\end{figure} 

\section{Method: A Structure Function Estimated Using Haar Wavelets}

The technique developed in this section was first used in \citet{2007ApJ...667.1024K} to study the time-structure of X-ray flares following {\it Swift}~GRBs.
It is applicable to a broad range of time-series.
Consider a time-series of length $N$ that can be regarded as one portion of one realization from the stochastic process $\{X_t, t = 0,\pm1, ...,T\}$ (considering a unit sampling interval).  Let
\be
\bar{X}_t (\tau) = \frac{1}{\tau} \sum_{n=0}^{\tau-1}{X_{t-n}} 
\ee
represent the sample average of $\tau$ consecutive observations, the latest one of which is $X_t$. The \citet{allan1966statistics} variance at scale $\tau$ is denoted by $\sigma_X^2 (\tau)$ and is defined to be half the mean square difference between adjacent non-overlapping $\bar{X_t} (\tau)$'s; i.e.,
\be
\sigma_X^2 (\tau) = \frac{1}{2} \left\langle [ \bar{X_t} (\tau) - \bar{X}_{t-\tau} (\tau) ]^2 \right\rangle 
\label{Eq:taylor}
\ee
Note that the \textit{Allan} variance at scale $\tau$ is a measure of the extent to which averages over length $\tau$ change from one time period of length $\tau$ to the next \citep{percival2006wavelet}. Comparing Equation \ref{Eq:taylor} to Equation \ref{Eq:sf}, we see that the \textit{Allan} variance is related to the SF of the smoothed signal $\bar{X_t}$.

In order to see how Haar wavelets can be used to estimate the SF (see, Figure \ref{fig:ep_pred2}), we will now relate the Haar wavelet coefficients to the averages calculated in determining the \textit{Allan} variance.  Consider the discrete Haar wavelet transform \citep{percival2006wavelet} of the time-series $X_1 , . . . , X_N$, where we assume that the sample size $N$ is a power of 2 so that $N = 2^q$ ($q > 0$). By definition, this transformation consists of $N - 1$ ``detail'' coefficients and one ``smooth'' coefficient $s_1 = \sum_{t=1}^{N}{\frac{X_t}{N}}$. The detail coefficients $d_{j,k}$ are defined for scales $k = 1, 2, 4, 8, . . . , N/2$ and - within the $k$th scale - for indices $j = 1, 2, 3, . . . , N/2k$ as
\be
d_{j,k} \equiv \frac{1}{\sqrt{2k}} \left [ \sum_{n=0}^{k-1}{X_{2 jk-n}} - \sum_{n=0}^{k-1}{X_{2 jk-k-n}} \right ]
\label{Eq:coef}
\ee

We can now state the relationship between the \textit{Allan} variance and the Haar wavelet coefficients $d_{j,k}$. Using Equations \ref{Eq:coef}, we have
\be
d_{j,k} = (\frac{k}{2})^{(1/2)} \left [  \bar{X}_{2jk} (k) - \bar{X}_{2jk-k} (k) \right ]
\ee
Under the assumption that $E\{d_{j,k}\} = 0$ so that the variance of $d_{j,k}$ is equal to $E\{d^2_{j,k}\}$, an average of the wavelet coefficients squared on scale $k$ provides a natural estimator for the {\it Alan} variance:
\begin{align}
\rm{var}\{d_{j,k}\} = \langle d_{j,k}^2\rangle = \frac{k}{2} \left\langle \left [  \bar{X}_{2jk} (k) - \bar{X}_{2jk-k} (k) \right ]^2 \right \rangle \notag \\
= \tau \sigma_X^2 (\tau)
\label{Eq:allan_haar}
\end{align}
We will redefine our wavelet coefficients (Equation \ref{Eq:coef}) by dividing out another factor of $\sqrt{k/2}$, eliminating the $k$ dependence in Equation \ref{Eq:allan_haar}.

\subsection{Data Analysis and Haar-SF Implementation}

In this section, we further develop our algorithm and discuss its application to GRB data captured by NASA's {\it Swift}~satellite \citep{2004ApJ...611.1005G}.
Our automated pipeline at Arizona State University is used to download the {\it Swift}~data in near real time from the {\it Swift}~Archive\footnote{ftp://legacy.gsfc.nasa.gov/swift/data} and quicklook site.  We use the calibration files from the 2008-12-17 BAT database release.  We establish the energy scale and mask weighting for the BAT event mode data by running the {\tt bateconvert} and {\tt batmaskwtevt} tasks from the HEASoft 6.12 software release\footnote{http://swift.gsfc.nasa.gov/docs/software/lheasoft/download.html}.  The mask weighting yields background-subtracted light curves. Light curves are extracted in the 15--350 keV band using the {\tt batbinevt} tool with 100 $\mu$s time bins, applying a uniform random deviate on the same timescale to undo artifacts associated with the data capture\footnote{See, http://swift.gsfc.nasa.gov/docs/swift/analysis/bat\_digest.html}. The burst duration intervals are determined automatically as described in \citet{2007ApJ...671..656B}.

Next, we group together adjacent time bins in the light curve until each composite bin has a fixed SNR of 5, dividing by the exposure time contained with each composite bin to produce the count rate versus time.  We then apply our analysis to the natural log of the binned light curve so that the error per light curve bin is approximately constant.  For homoscedastic errors -- as we now have -- the orthogonality properties of the Haar wavelets lead to approximate statistical independence of the wavelet coefficients \citep[see, e.g.,][]{percival2006wavelet}. We find that working with binned data with approximately constant errors from point to point leads to the most stable SF estimates.  

Producing a scaleogram using the logarithm of the count rate can be interpreted as yielding the average {\it fractional}~change in the signal versus timescale.  We believe such a measure allows for more physical insight into the emission mechanism than a measure of absolute change versus timescale.  We note that the time binning will no longer be uniform.  This is not a problem for the analysis, provided we propagate the true time difference associated with each wavelength coefficient through the analysis.  We call the resulting scaleogram the Haar wavelet structure function, and we will denote it as $\sigma_{X,\Delta t}$ below.  To improve statistics, we calculate $\sigma_{X,\Delta t}$ using not just one discrete Haar transform, but averaging $\sigma_{X,\Delta t}$ over the $N$ transforms resulting from cyclic permutations of the data of length $N$ (i.e., the non-decimated Haar transformation).

\subsection{A Sample Burst: The ``Naked-Eye'' GRB 080319B}
\label{sec:naked}

\begin{figure}
\centering
\includegraphics[width=.42\textwidth]{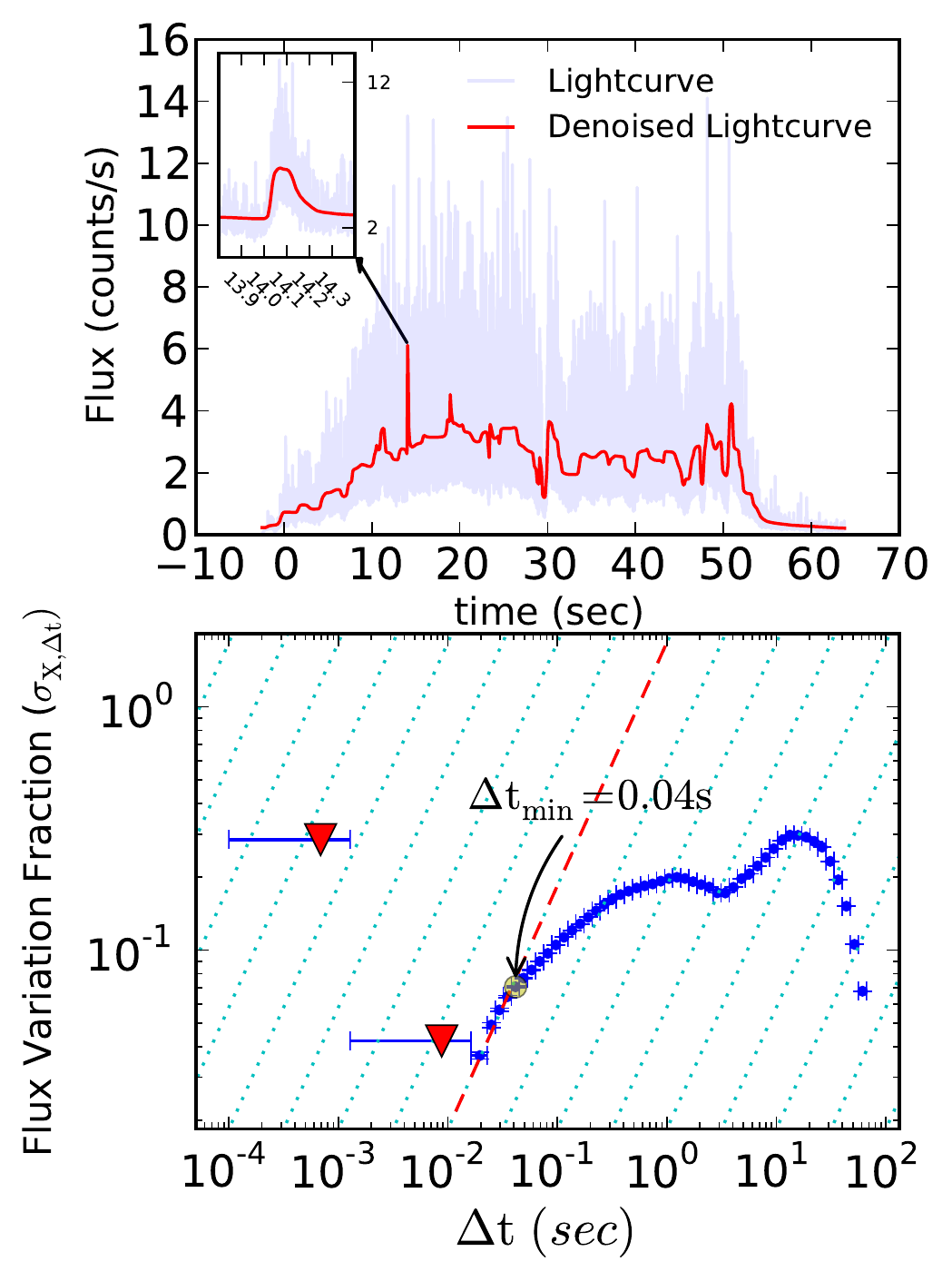} 
\caption{\small 
(Top) \textit{Swift} BAT light curve (15--350 keV band) of the ``Naked Eye'' GRB~080319B. Overplotted is a denoised version of the light curve (following \citet{1997ApJ...483..340K}, see also \citet{2002A&A...385..377Q}), highlighting the true signal variation that would be observed were there no Poisson error on the measurement.  (Bottom) The Haar wavelet scaleogram $\sigma_{X,\Delta t}$ vs. timescale $\Delta t $ for GRB~080319B.  We show only $3 \sigma$ excesses over the power associated with Poisson fluctuations and report lower values as $3 \sigma$ upper limits using red triangles.   We derive a minimum timescale of $40 \pm 10$ ms (Section \ref{sec:naked}), corresponding to the shortest timescale at which $\sigma_{X,\Delta t}$ departs from $\sigma_{X,\Delta t} \propto \Delta t$.  The corresponding light curve structure can be seen clearly in the top panel inset.}
\label{fig:pick_burst}
\end{figure}

We now implement the Haar structure function on a real GRB light curve.
The prompt BAT Gamma-ray light curve for GRB~080319B \citep[see, e.g.,][and references therein]{2009ApJ...691..723B} and our derived $\sigma_{X,\Delta t}$ curve are shown in Figure \ref{fig:pick_burst}.   To guide the eye, several lines of constant $\sigma_{X,\Delta t} \propto \Delta t$ are also plotted.   The expected measurement error has been subtracted away, leaving only the fractional flux variation expected for the GRB.  On timescales where this net variation is greater than zero at the $3\sigma$ confidence level, we plot data points.  Otherwise we plot upper limits.   

Although there is excess signal present on timescales as short as 20 ms in Figure \ref{fig:pick_burst} (bottom), these timescales correspond to a region of the plot where $\sigma_{X,\Delta t} \propto \Delta t$ and should be interpreted as being due to temporally-smooth variations in a signal which is varying in an unsmooth fashion on longer timescales.  The $\sigma_{X,\Delta t}$ points pull away significantly (2$\sigma$ level from a $\Delta \chi^2$ test) from the $\sigma_{X,\Delta t} \propto \Delta t$ curve at $\Delta t_{\rm min} = 40 \pm 10$ ms.  This is the timescale of interest, describing the minimum variability time for uncorrelated variations in the GRB.

Beyond this timescale, $\sigma_{X,\Delta t}$ is flatter than $\sigma_{X,\Delta t} \propto \Delta t$, indicating the presence of pulses with typical durations on these timescales.  On a timescale of about 1 s, the $\sigma_{X,\Delta t}$ begins turning over due to a lack of signal variation between this timescale and the timescale (tens of seconds) describing the emission envelope.  We are not concerned here with those longer timescale structures, although we do note that $\sigma_{X,\Delta t}$ provides a rich, aggregate description of this temporal activity.

In the inset to Figure \ref{fig:pick_burst} (top), we show a zoom-in on the narrowest time structure present in the signal.  It can be seen that the approximate rise-time of this pulse corresponds nicely to our derived minimum timescale.

\subsection{Simulated GRBs}

\begin{figure}
\begin{center}
\includegraphics[width=.45\textwidth]{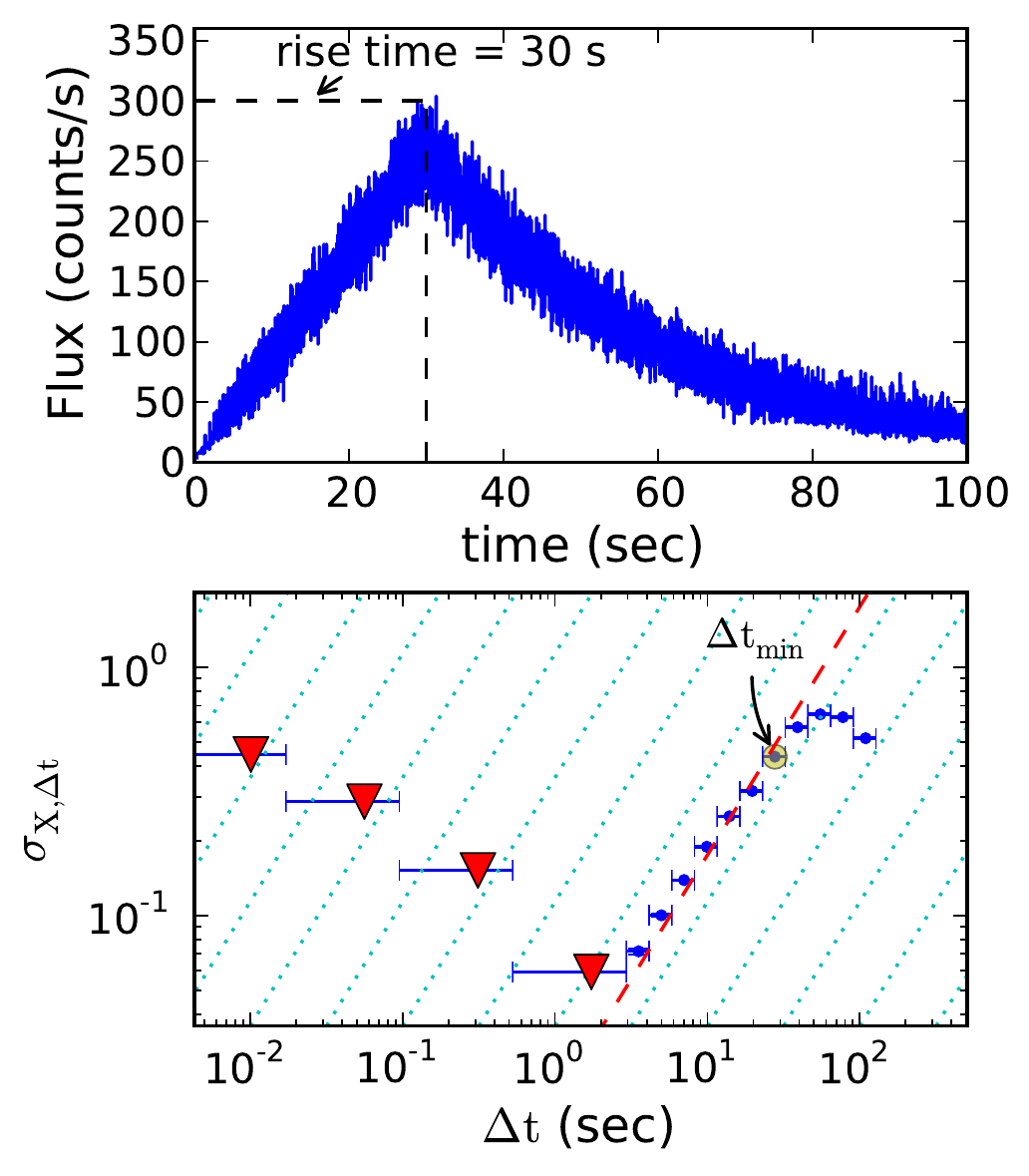}  
\caption{\small 
A simulated FRED pulse with a rise time of 30 s (Top) and the corresponding Haar wavelet scaleogram $\sigma_{X,\Delta t}$ vs. timescale $\Delta t$ (Bottom). The expected Poisson noise level has been subtracted.
The scaleogram rises linearly on short timescales corresponding to regions of the light curve where the signal is varying smoothly, falling away from the $\sigma_{X,\Delta t}\propto \Delta t$ trend
on a timescale ($\approx 30$ s).  This minimum timescale can be robustly measured, provided there is sufficient SNR for the linear region preceding it to be well identified.}
\label{fig:simburst_30s}
\end{center}
\end{figure}

The above example demonstrates that $\sigma_{X,\Delta t}$ can be used to extract a minimum timescale from a bright GRB with a rich temporal profile.
How robust would the recovery of this timescale be for fainter GRBs?  We examine this question using simulated pulses.  Figure \ref{fig:simburst_30s}
displays a simulated GRB, consisting of a single pulse with FRED profile.
The minimum timescale -- corresponding to the rise of the pulse -- can be correctly identified because there is sufficient SNR to identify
the $\sigma_{X,\Delta t}\propto \Delta t$ region of the plot preceding that timescale.

\begin{figure}
\begin{center}
\includegraphics[width=.47\textwidth]{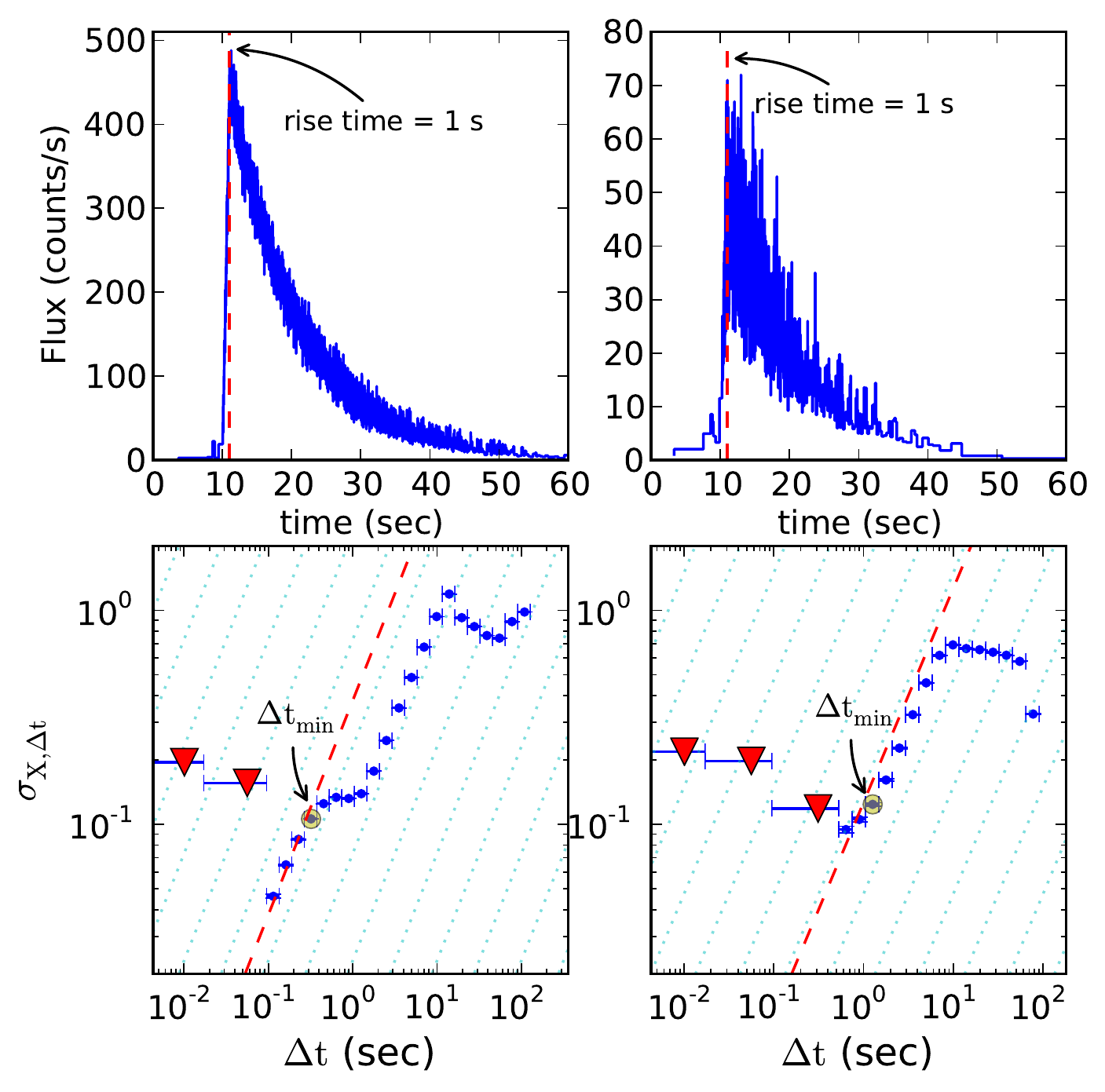}  
\caption{\small 
Simulated FRED pulses with a rise time of 1 s (Top). The left pulse is an order of magnitude brighter than the right pulse.
The bottom panels show the corresponding Haar wavelet scaleogram $\sigma_{X,\Delta t}$ vs. timescale $\Delta t $. The expected level for Poisson noise has been subtracted.
It is clear that fine-time structure can be missed in the low SNR limit; however the $\sigma_{X,\Delta t}$ measurements remain consistent.}
\label{fig:simburst_4}
\end{center}
\end{figure}

We now consider a pulse with markedly different rise-time $T_{\rm rise}$ and total duration $T_{\rm tot}$, over a range of possible brightness.
Figure \ref{fig:simburst_4} shows a simulated GRB (single pulse) with $T_{\rm rise}/T_{\rm tot} \sim 1/100$.  The left panel shows a pulse which is an order of magnitude brighter than that in the right panel.
The bottom panels of Figure \ref{fig:simburst_4} display the corresponding fractional flux variation $\sigma_{X,\Delta t}$ as a function of timescale $\Delta t$.
The derived minimum timescale in case of brighter light curve is close to the rise-time of 1 s.

When we decrease the pulse SNR, the linear rise phase spans less of the plot and is harder to identify (Figure \ref{fig:simburst_4} (right)); it becomes more difficult to identify the minimum timescale.  In this example, $\Delta t_{\rm min}$ is still identified correctly for 10 times lower SNR.  We note that the y-axis levels ($\sigma_{X,\Delta t}$) in the low and high SNR plots are consistent:  we infer the correct fractional signal power at each $\Delta t$ in each case.

With further decreasing SNR levels, the linear rise phase in $\sigma_{X,\Delta t}$ due to the pulse rise will become absent.  If the first non-zero $\sigma_{X,\Delta t}$ values lies on a line flatter than linear, the associated $\Delta t_{\rm min}$ value must be regarded as an upper limit.  We will follow this convention below: if a linear rise phase on the shortest timescales cannot be confirmed, $\Delta t_{\rm min}$ will be taken as an upper limit.

At very low SNR levels, our analysis will tend to miss the linear rise phase in $\sigma_{X,\Delta t}$ associated with the pulse rise time and will instead identify the pulse decay time (or total GRB duration) as the minimum timescale.  Because actual GRB pulses tend to be asymmetric like this simulated pulse, we expect our $\sigma_{X,\Delta t}$ analysis to correctly identify true minimum timescale if is within an order of magnitude (or so) of the lowest measurable timescale.  Care will have to be taken when considering faint GRBs with $\sigma_{X,\Delta t}$ of order unity and with $\Delta t_{\rm min}$ comparable to the event duration.

We conclude from this that the measurement of a linear rise phase in $\sigma_{X,\Delta t}$, followed by a flattening, will allow us to infer the presence of a characteristic timescale describing the transition from smooth (correlated) to unsmooth (uncorrelated) variability.   However, we will not generally be able to rule on the presence of uncorrelated variability on much shorter (i.e. factor 10-100) timescales.
We stress again that the shortest timescale exhibiting a net $\sigma_{X,\Delta t}$ over the Poisson level (called $\Delta t_{\rm snr}$ below),  is not a timescale with intrinsic meaning independent of the noise level.

\section{Discussion and Results}

\begin{figure*}
\begin{center}
\includegraphics[width=.9\textwidth]{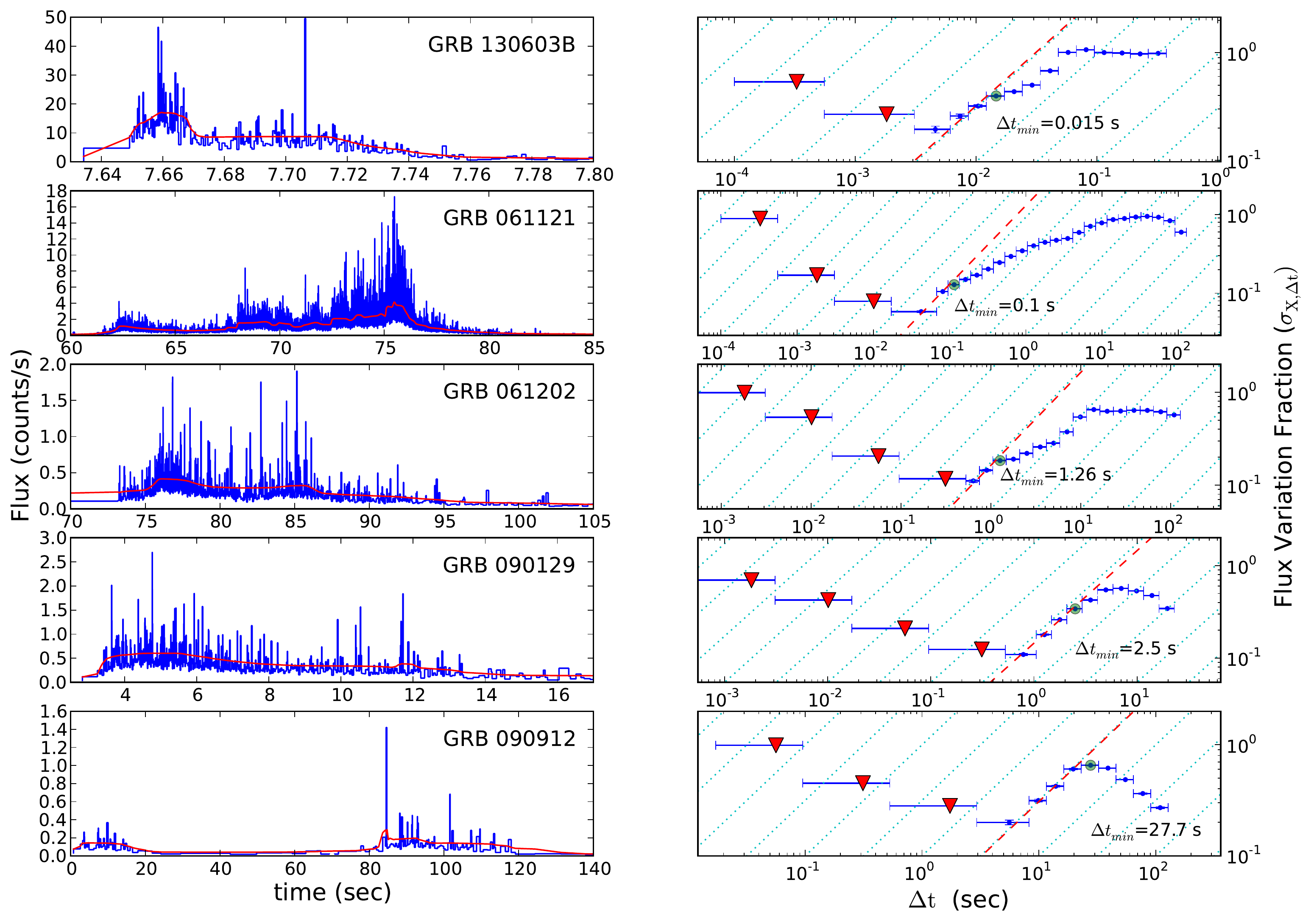} 
\caption{\small
A gallery of Haar scaleograms $\sigma_{X,\Delta t}$, spanning a range of minimum variability timescale $\Delta t_{\rm min}$. The left panel shows the light curve in blue, with a denoised red curve to guide the eye \citep[following][]{1997ApJ...483..340K}. The corresponding Haar scaleogram plot is shown in the right panel.  In each of these, the red dashed-line represents the temporally-smooth ($\sigma_{X,\Delta t} \propto \Delta t$) region and the green circle marks the extracted $\Delta t_{\rm min}$.}
\label{fig:gallery}
\end{center}
\end{figure*}

We analyze the \textit{Swift} data set up until October 27, 2013, which consists of 744 GRBs, 251 with measured redshifts.
We only consider those GRBs with total light curve ${\rm SNR} \geq 10$, leaving 517 GRBs. Of these, we are able to confirm the presence of a linear rise phase in $\sigma_{X,\Delta t}$ on short timescales for 281 GRBs.  We quote upper-limit values for the remainder.
Most (256) of the bursts in this compiled subsample are long-duration ($T_{\rm 90}>3$ s) GRBs.  In the compiled subsample, 98 GRBs in the compiled subsample have measured redshift.
The temporal specifications of all 517 GRBs discussed here are determined using fully-automatic software and are presented in Table 1.
Light curves and Haar scaleograms are shown for a subset of the GRBs in Figure \ref{fig:gallery}.

\begin{figure}
\begin{center}
\includegraphics[width=.47\textwidth]{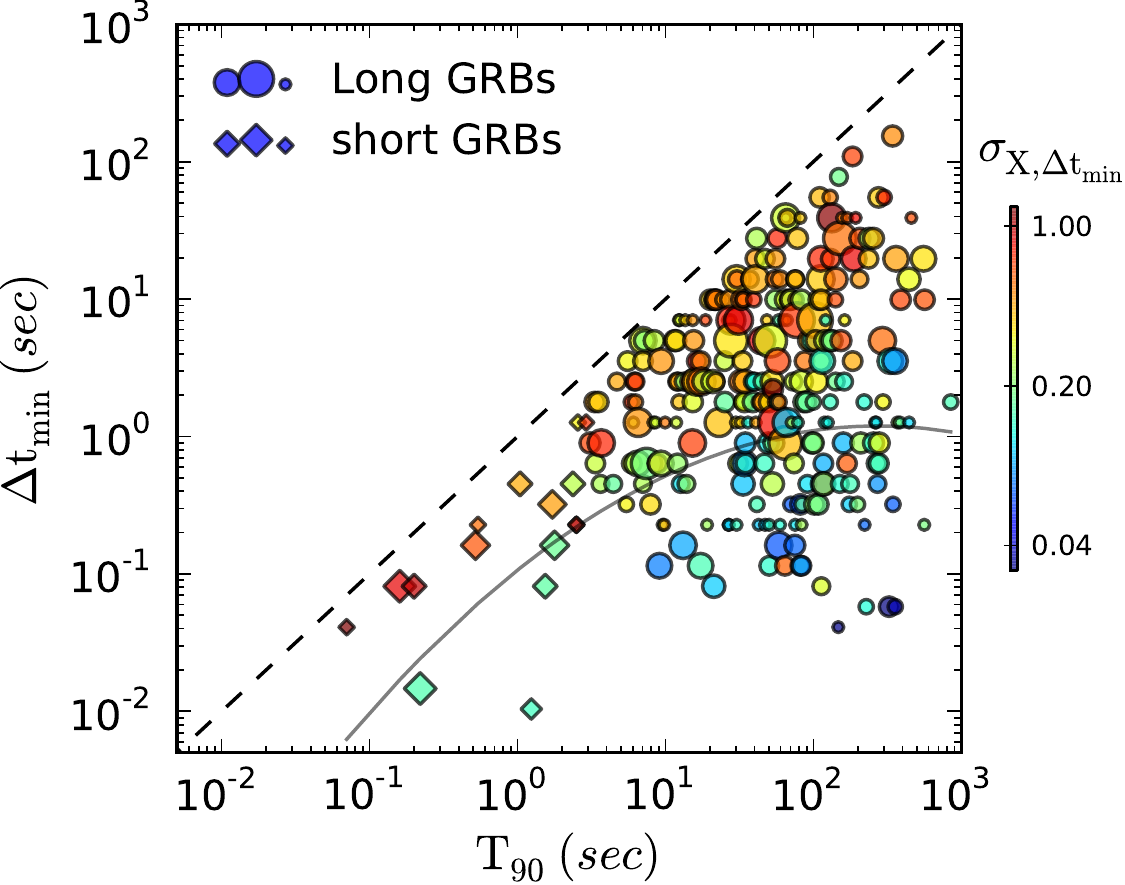}  
\caption{\small 
Our GRB minimum time $\Delta t_{\rm min}$ plotted versus the GRB $T_{\rm 90}$ duration.  Circles (diamonds) represent long-duration (short-duration) GRBs.
 The point colors represent the fractional flux variation level ($\sigma_{X,\Delta t_{\rm min}}$) at $\Delta t_{\rm min}$.  Also plotted as a curved line is the typical minimum observable $\Delta t_{\rm min}$ as a function of $T_{\rm 90}$.  The symbol sizes are set by the ratio of $\Delta t_{\rm min}$ to the actual minimum observable time ($\Delta t_{\rm snr}$) for each GRB.}
\label{fig:T90vsTmin}
\end{center}
\end{figure}

Figure \ref{fig:T90vsTmin} displays our minimum variability timescale, $\Delta t_{ \rm min}$, versus the GRB duration, $T_{\rm 90}$. The short and long-duration GRBs are shown with diamond and circle symbols, respectively.  In this plot the relative size of symbols is set by the ratio between minimum variability and SNR timescale ($\Delta t_{\rm min} / \Delta t_{\rm snr}$).  As described above, $\Delta t_{\rm snr}$ represents the first statistically significant timescale in the Haar wavelet scaleogram.  The color of the points in Figure \ref{fig:T90vsTmin} corresponds to the fractional flux variation level ($\sigma_{X,\Delta t}$) at $\Delta t_{\rm min}$.   A curved black line is also plotted to show a typical value for the minimum observable time ($\Delta t_{\rm snr}$) versus the GRB duration, $T_{\rm 90}$ \citep[from][]{2007ApJ...671..656B,2010ApJ...711..495B,2013AstRv...8a.103B}.

We first note from the colors in Figure \ref{fig:T90vsTmin}, that GRBs with $\Delta t_{\rm min}$ close to $T_{\rm 90}$ tend to have flux variation fractions of order unity.  These are bursts with simple, single-pulse time profiles.  As can be seen from the range of point sizes in Figure \ref{fig:T90vsTmin}, most are not simply low SNR events where fine time structure cannot be observed.  Also, we see that there are GRBs with both high and low SNR which have complex time-series ($\Delta t_{\rm min} \ll T_{\rm 90}$).  Based again on the point colors, this short-timescale variation tends occur at a small fractional level in the signal ($\sigma_{X,{\Delta t}} \approx $ 1--10 \%), at least for the long-duration GRBs.

From a Kendall's $\tau$-test \citep{kendall1938new}, we find only marginal evidence that $\Delta t_{\rm min}$ and $T_{\rm 90}$ are correlated ($\tau_k = 0.38$, $1.5\sigma$ above zero). The correlation strength is marginally stronger if we utilize the robust duration estimate $T_{\rm R45}$ \citep{reichart01} in place of $T_{\rm 90}$: $\tau_k = 0.6$ ($2.4\sigma$). The $\Delta t_{\rm min}$ values in Figure \ref{fig:T90vsTmin} are bound from above by $T_{\rm 90}$, and they do not strongly correlate with $T_{\rm 90}$ within the allowed region of the plot.  Recently, \citet{2013MNRAS.432..857M} have studied faint Fermi GBM GRBs and do find evidence for a correlation.  We can reconcile our conclusions by identifying low SNR as the driving force in any apparent correlation. If we perform a truncated Kendall's $\tau$ test which only compares GRBs above one-another's threshold \citep{2002ApJ...565..182L}, the correlation strength drops precipitously ($\tau_k = 0.14$, $0.5\sigma$  for $T_{\rm 90}$; $\tau_k = 0.4$, $1.5\sigma$ for $T_{\rm R45}$). We, therefore, believe there is no strong evidence supporting a real correlation between $\Delta t_{\rm min}$ and $T_{\rm 90}$.

\begin{figure*}
\begin{center}
\includegraphics[width=.95\textwidth,height=2.4in]{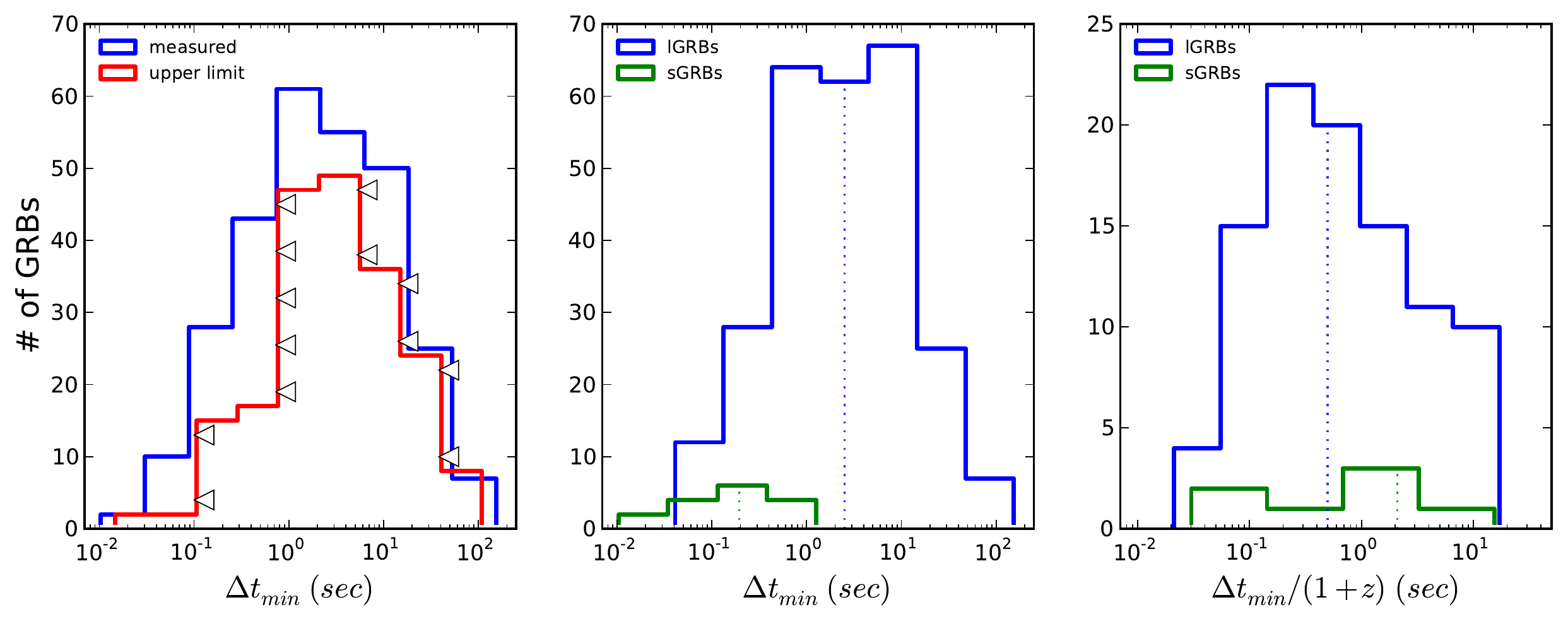}  
\caption{\small 
The histograms of $\Delta t_{\rm min}$ with reliable measurement (blue) and for GRBs allowing for upper limits only (red).  
In observer frame (middle panel) the median minimum timescale for long-duration GRBs is: $\Delta t_{\rm min} = 2.5$ s, and for short-duration GRBs is $\Delta t_{\rm min} = 0.2$ s.
The same quantities in source frame (right panel) are: $\Delta t_{\rm min} = 0.5$ s and $\Delta t_{\rm min} = 2.1$ s.
}
\label{fig:hist}
\end{center}
\end{figure*}

Figure \ref{fig:hist} (left) shows histograms for the {\it Swift}~GRBs with reliable $\Delta t_{\rm min}$ (in blue) measurement and also the GRBs for which only upper limits on $\Delta t_{\rm min}$ could be derived (in red).
The distributions have consistent mean values. (We discuss discrepancies between the tails of the distributions below.)
We find a median minimum timescale for long-duration (short-duration) GRBs in the observer frame of 2.5 s (0.2 s).  In the source frame, we find a median minimum timescale for long-duration (short-duration) GRBs of 0.5 s (2.1 s).

From Figure \ref{fig:hist}, we observe that the $\Delta t_{\rm min}$ distribution of long-duration GRBs is displaced from that of short-duration GRBs ($8\sigma$, $t$-test).
 This finding is consistent with the findings of \citet[][see also \citet{2001grba.conf...40N}]{2013MNRAS.432..857M}, but not those of \citet{2000ApJ...537..264W} who find statistically indistinguishable distribution centers.  We do note that the distribution centers appear to be consistent when viewed in the source frame (Figure \ref{fig:hist} (right)), although the number of short-duration GRBs with redshift is low.
The reason for the observer frame discrepancy is likely the fact that short-duration GRBs tend to be detected only at low-redshift, unlike long-durations GRBs which span a broad range of redshifts.

Examining the dispersion in $\log(\Delta t_{\rm min})$ values, shows no strong evidence for dissimilar values for the long and short-duration samples ($<1.5\sigma$, $F$-test).   This finding is fully consistent with \citet{2013MNRAS.432..857M}, where it was also found (using a larger sample of short-duration GRBs) that the two histograms are quite broad and very similar in dispersion.


\begin{figure}
\centering
\includegraphics[width=.50\textwidth]{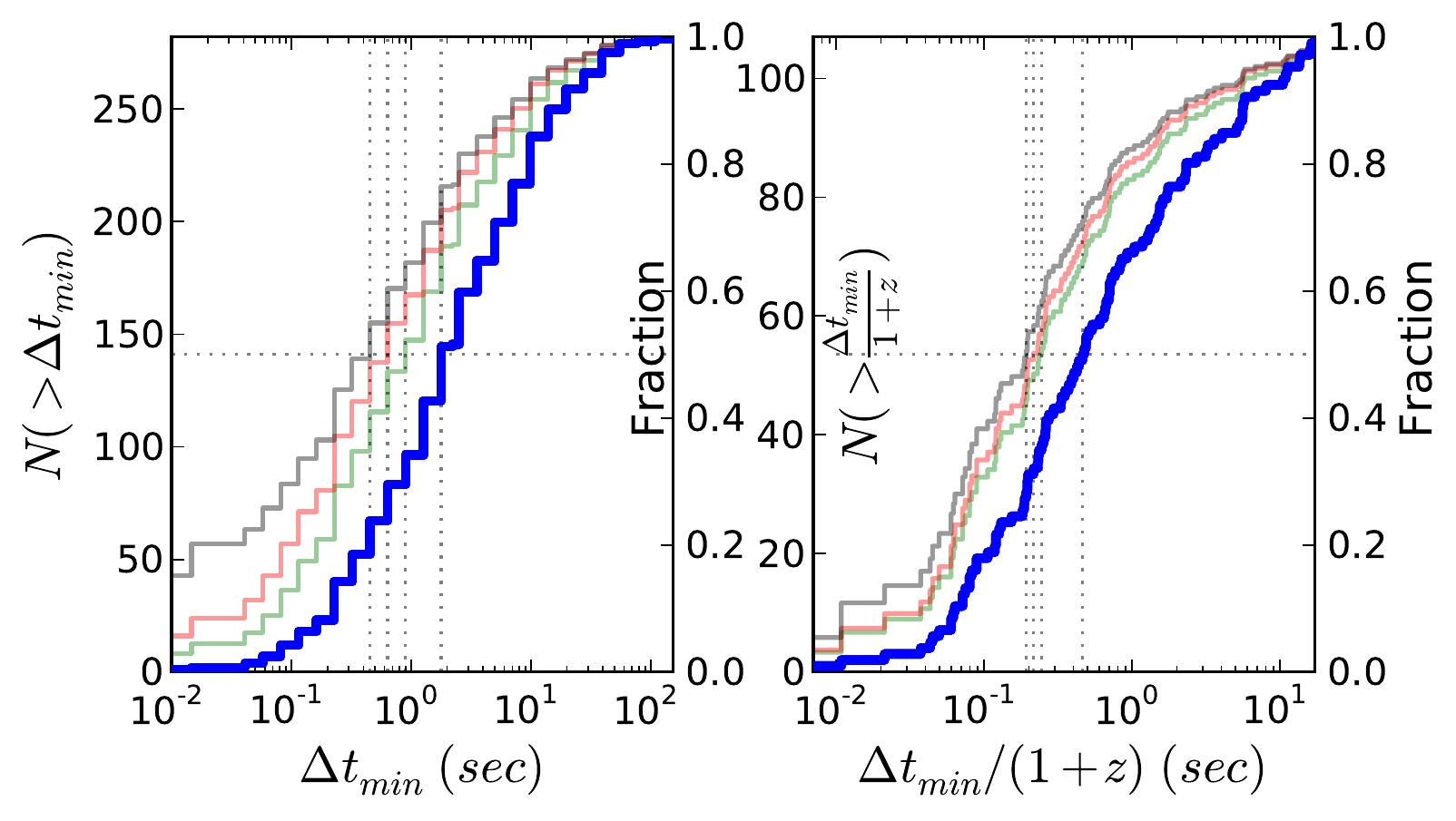}  
\caption{\small
Cumulative distributions in the observer frame (left) and source frame (right) for $\Delta t_{\rm min}$ for GRBs with well measured values (blue).
The Kaplan-Meier estimator \citep{1985ApJ...293..192F} is used to also include GRBs with upper limits on $\Delta t_{\rm min}$ (green, red, and gray curves).
The median minimum timescale for GRBs with long and short-durations in the observer frame is: $\Delta t_{\rm min} = 1.8$ s, and in source frame is: $\Delta t_{\rm min} = 0.5$ s.}
\label{fig:tick}
\end{figure}

We have a large sample of GRBs for which no $\Delta t_{\rm min}$ could be calculated or for which only upper limits on $\Delta t_{\rm min}$ were obtainable.  To account for the relative frequencies of such GRBs, we employ a survival analysis \citep[see, e.g.,][]{1985ApJ...293..192F}.
Figure \ref{fig:tick} displays the Kaplan-Meier estimator curves, which combine the detections and upper limits. There are three thin curves (green, red, and gray) in each panel of Figure \ref{fig:tick}.  The green curve is calculated  including the derived upper limits for $\Delta t_{\rm min}$ when SF fitting was possible.  The red curve includes these as well as $\Delta t_{\rm min}$ estimates for the remaining GRBs, where we take $T_{\rm 90}$ as the limiting value of $\Delta t_{\rm min}$ when no SF fitting was possible. The gray curve is similar to the red curve, with $T_{\rm R45}$ used in place of $T_{\rm 90}$ for the limiting value of $\Delta t_{\rm min}$. The median minimum timescale for GRBs (long and short-duration) in the observer frame is $\Delta t_{\rm min} = 1.8$ s. In the source frame, the median is $\Delta t_{\rm min} = 0.5$ s.  The survival analysis does not strongly affect these median values.

\citet{2000ApJ...537..264W}, using BATSE data, report that most GRBs appear to exhibit millisecond variability. Claims have also been made for the presence of sub-millisecond variability \citep{1992Natur.359..217B}, and even micro-second variability (\citet{1989Ap&SS.155..141M}, but see \citet{1993ApJ...404..673S,1997ApJ...491..720D}). In contrast, we find that only 0.4\% of {\it Swift}~BAT GRBs with well-measured $\Delta t_{\rm min}$ have $\Delta t_{\rm min}< 10$ ms (observer frame). If we include all {\it Swift}~GRBs using the survival analysis, we still find a fraction below 6\% using $T_{\rm 90}$ and below 15\% using $T_{\rm R45}$.  In the source frame, the numbers are 1\% (well-measured) and 4\% (all, using $T_{\rm 90}$) and 5\% (all, using $T_{\rm R45}$). 
Of 517 bursts where 1 ms variability could have been measured, none show such short-timescale variability. We, conclude that millisecond variability may be quite rare.

\subsection{Evidence for Time-Dilation?}

Given that we can derive a robust GRB minimum timescale and that redshifts have been measured for many of our GRBs, it is interesting to test whether these quantities are correlated.  As GRBs are present over a very broad redshift range the signature of time-dilation ought to be present in GRB time-series.  However, finding such a signal has remained elusive (\citet{1994ApJ...424..540N}; \citet{2013ApJ...765..116K}, but see, e.g., \citet{2013ApJ...778L..11Z}).

\begin{figure}
\begin{center}
\includegraphics[width=.47\textwidth]{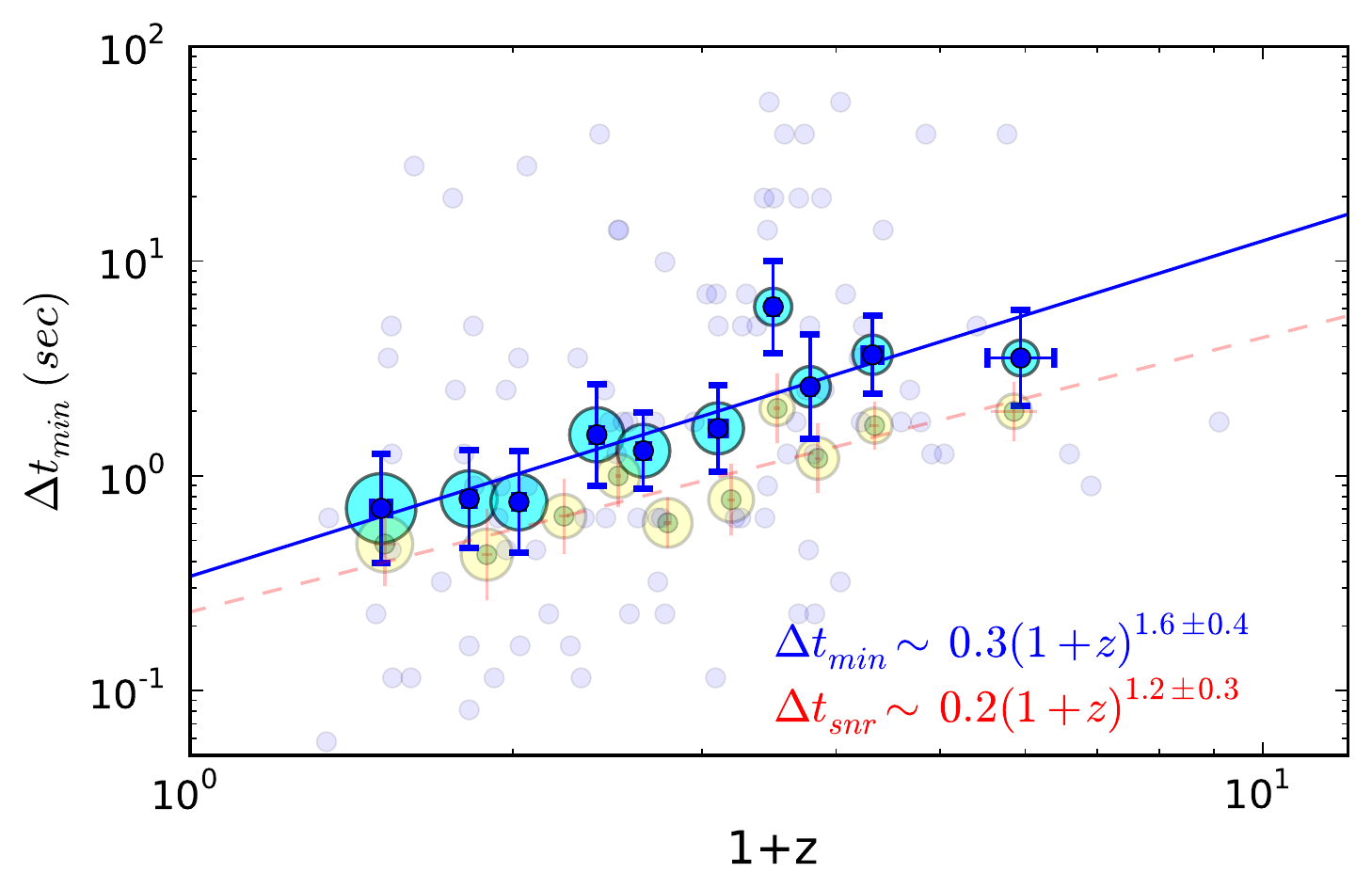}  
\caption{\small 
Minimum variability timescale in the observer frame versus redshift $z$.
The blue crosses show geometric averages of $\Delta t_{\rm min}$ with 10 bursts in each bin and red crosses show geometric averages of $\Delta t_{\rm snr}$ with 18 bursts in each bin.
Cyan and yellow circles correspond to the average of SNR of bursts in each bin. 
The faint blue circles show all GRBs with measured $\Delta t_{\rm min}$ and known $z$.}
\label{fig:sbolite}
\end{center}
\end{figure}

In Figure \ref{fig:sbolite}, we plot $\Delta t_{\rm min}$ as a function of redshift.  Redshift values are taken from \citet[and references therein]{2007ApJ...671..656B,2010ApJ...711..495B}.  The blue crosses in Figure \ref{fig:sbolite} correspond to geometric averages for sets of 10 bursts of similar redshift.  The unbinned data are plotted in the background.  We find that the binned data can be well-fitted by a line $\Delta t_{\rm min} = 0.3 (1+z)^{1.6\pm0.4}$, possibly indicating the presence of time-dilation, although with a larger best-fit power-law index than would naively be expected. The effects of widening pulse width with decreasing observed energy bandpass (as a result of increasing redshift) are expected to play a role in increasing our observed index relative to that predicted from cosmological time dilation only \citep[see,][]{1995ApJ...448L.101F,1996ApJ...459..393N}.

The purity of this relation comes into question, however, when we perform a similar fit to the minimum observable timescale.  We find that this quantity also correlates with redshift, as $\Delta t_{\rm snr} = 0.2 (1+z)^{1.2\pm0.3}$.  This must be the result of selection effects: more distant GRBs are fainter, thus permitting measurement of only long variability timescales.  The power-law indices of the two fits are statistically consistent (1-$\sigma$ level).  For the unbinned data, we find $\tau_k = 0.24$ ($3.6\sigma$), but this drops to $\tau_k = -0.02$ accounting for the limits.  Given the clear role the threshold plays in defining the correlation strength, we cannot be confident that time-dilation is being uniquely measured by $\Delta t_{\rm min}$.

\section{Conclusions}

Using a technique based on Haar wavelets, we have studied the temporal properties of a sample of GRB hard X-ray, prompt-emission light curves captured by the BAT instrument on {\it Swift} prior to October 27, 2013.
Our approach averages over the time-series captured for a given GRB, providing robust measures of minimum variability timescales.  

In contrast to previous studies \citep{2000ApJ...537..264W,2013MNRAS.432..857M,2013arXiv1307.7618B}, which simply define the minimum timescale in reference to the measurement noise floor, our approach identifies the signature of temporally-smooth features in the wavelet scaleogram and then additionally identifies a break in the scaleogram on longer timescales as signature of a true, temporally-unsmooth light curve feature or features. We find that this timescale ($\Delta t_{\rm min}$) tends to correspond to the rise-time of the narrowest GRB pulse \citep[see, also,][]{2013arXiv1307.7618B}.

We find a median minimum timescale for long-duration GRBs in the source (observer) frame of $\Delta t_{\rm min}=0.5$ s ($\Delta t_{\rm min}=2.5$ s).  A consistent value in the source-frame for short-duration GRBs may indicate a common central engine.

We find that very few -- at most 15\% (5\%) in the observer frame (source frame) -- of {\it Swift}~GRBs can have minimum timescales below 10 ms.  Our timescales are thus considerably longer than the millisecond variability timescales found by \citet{2000ApJ...537..264W} to be common in bright BATSE GRBs.  Partial explanation for this discrepancy must come from the fact that {\it Swift}~BAT operates in a lower photon energy range than BATSE, and GRB pulses are known to be more narrow in higher energy bandpasses \citep[e.g.,][]{1996ApJ...459..393N,1995ApJ...448L.101F,1996ApJ...473..998F}.  Nonetheless, we note that the variability found in \citet{2000ApJ...537..264W} is not linked to the presence of discernible features in a given light curve (e.g., the pulse rise-time, which it is actually stated to be considerably less than).  Given our new distinction between a \citet{2000ApJ...537..264W} type timescale (which we call $\Delta t_{\rm snr}$) -- the minimum possible observable $\Delta t_{\rm min}$ for a GRB of given brightness and not necessarily the true $\Delta t_{\rm min}$ -- it is natural to expect that \citet{2000ApJ...537..264W} have underestimated their minimum timescales. 
We note that our minimum timescales are broadly consistent with those found in pulse-fitting studies \citep[e.g.,][]{1995ApJ...448L.101F,1996ApJ...459..393N}.

\subsection{Constraints on the Fireball Model}

The standard fireball model postulates the release of a large amount of energy by a central engine into a concentrated volume \citep{1978MNRAS.183..359C,2004RvMP...76.1143P}, which causes the resulting outflow to expand and quickly become relativistic \citep{1986ApJ...308L..43P}. These relativistic expanding shells -- with different Lorentz factors -- in general collide, resulting in Gamma-ray flares and potentially rich temporal structure. Our extracted timescales ($\Delta t_{\rm min}$) should provide a diagnostic on the central engine power and its evolution. The size of the central engine is limited to $R < c \, \Delta t_{\rm min}$, which for the smallest minimum variability timescale derived above \citep[$\sim 10$ ms; see also,][]{2002MNRAS.330..920N} is $R < 3 \times 10^3$ km.  Typical $\Delta t_{\rm min}$ values from above lead to $R < 2 \times 10^5$ km.  For the first time, due to a large sample of GRBs with measured redshift, we are able to perform these calculations in the source frame.
	
In the source frame, we are not able to confirm that the minimum variability timescale of short-duration GRBs is substantially shorter than that of long-duration.  Hence, we cannot demonstrate that short-duration GRBs have a more condensed central engine than the former \citep[see,][]{2013arXiv1307.7618B}.

We can derive additional constraints on the GRB emission region,
following the discussion in \citet{2000ApJ...537..264W}.

In the ``external shock'' picture, shells of material produced by the GRB impact material in the external medium.
The physical dimension of clouds and their patchiness -- in the direction perpendicular to the expansion of the shell -- is constrained by $\Delta t_{\rm min}$. If we assume a single shell expanding at very close to light speed, the arrival time for photons from the shell will be calculated as the summation of travel time of the shell to the radius of impact and the travel time of the Gamma-rays to Earth. Photons from off-axis regions of the relativistic expanding shell experience a purely geometrical delay compared with photons from on-axis regions \citep{2004RvMP...76.1143P} reaching the observer. The observed delay depends only on the radius of the shell $R$ at the time of impact with a cloud in the external medium and the angular radius of the Gamma-ray emission region as subtended from the burst site ($\Delta \Theta$). 

\citet{2000ApJ...537..264W} report millisecond variability superposed on pulses of significantly longer rise-times. High cloud patchiness can potentially explain this modulation, implying $\Delta \Theta < \Delta t_{\rm min}/(2\Gamma\, T_{\rm rise}) < 0.0002$ radians, where $\Gamma$ is the bulk Lorentz factor.  Only $\sim 5 \times 10^{-3}$ of the emitting shell is active at a given time \citep[also,][]{1999ApJ...512..683F}.  However, for the bursts in our sample, with typical minimum variability timescale $\sim 0.1 \, s$,  $T_{\rm rise} >  1 \, s$, and assuming $\Gamma > 100$, we find that $\Delta \Theta < 5 \times 10^{-3}$ radians which is comparable with the typical surface filling factor \citep{1999ApJ...512..683F}.  

In terms of the external shock scenario, the extracted $\Delta t_{\rm min}$ can circumscribe the size scale of the impacted cloud along the line of sight. For a thin shell, the Gamma-ray radiation will start when the relativistic shell hits the inner boundary of the cloud with the peak flux produced as the shell reaches the densest region or center of the cloud. The size scale of the impacted cloud is limited  by $2 \Gamma^2 c\,\Delta t_{\rm min}$ since the shock is moving near light speed \citep{1996ApJ...473..998F}. For the smallest $\Delta t_{\rm min}$ found $\sim 10$ ms,  and assuming $\Gamma < 1000$, the cloud size must be smaller than $40$ AU.

In the ``internal shock'' scenario \citep[e.g.,][]{1994ApJ...430L..93R}, the relativistic expanding outflow released from a central engine is assumed to be variable, consisting of multiple shells of differing $\Gamma$. These shells propagate and expand adiabatically until a faster shell collides with a slower one, resulting in a measurable rise time. This rise time to an outside observer would appear as: $\Delta t_{r1} \approx \Delta R / 2c\Gamma_1^2$, where $\Delta R$ and $\Gamma_1$ are the thickness and resulting Lorentz factor of the merged shell \citep[e.g.,][]{2007ApJ...667.1024K}. Assuming the same scenario but for two other shells yields $\Delta t_{r2} \approx \Delta R / 2c\Gamma_2^2$. Writing $\Delta \Gamma = \Gamma_1 - \Gamma_2$, we have $\Delta \Gamma / \Gamma \approx 1/2 \, (\Delta t_{\rm min} / T_{\rm rise})$.  In \citet{2000ApJ...537..264W}, the ratios $\Delta t_{\rm min} / T_{\rm rise}$ were argued to be small, implying a narrow dispersion in $\Gamma$.  We, however, find $\Delta t_{\rm min} \sim T_{\rm rise}$, suggesting instead a broad range of possible Lorentz factors.

Finally, we find evidence that our minimum timescales correlate with redshift, possibly providing indication of cosmological time-dilation.  However, the measurement threshold also appears to correlate strongly with redshift.  This indicates that threshold effects likely dominate the apparent correlation and that the correlation may not be real.  It is possible that additional features present in the Haar scaleogram (slopes, breaks on longer timescales) -- which richly describe the full GRB light curve and not just the minimum timescale -- may yield correlations with intrinsic quantities like redshift.  We will study this further in future work.

\acknowledgments
We thank Owen Littlejohns, Judd Bowman, and Teresa Ashcraft for comments on the manuscript and for useful discussions. We also thank our referee, Jay Norris, for insightful comments that improved the manuscript.

%\clearpage
%\vspace{0.5in}
%\cite{*}

\clearpage

\vfill
\eject
\clearpage
\begin{center}
\begin{scriptsize}
\begin{longtable}{@{}l@{}c@{}c@{}c@{}c@{}c@{}c@{}c@{}c@{}c}

\caption[]{GRB Minimum Timescales} \label{tab:spec_table} \\

%This is the header for the first page of the table...
\hline \hline \\[2ex]
   GRB & \; $\Delta t_{\rm min}$ & \; $\Delta t_{\rm snr}$ & \; $T_{\rm 90}$ & \; $T_{\rm R45}$ & \; $\sigma_{X, t_{\rm min}}$ & \; $\sigma_{X, t_{\rm snr}}$ & \; $\chi^2/\nu$ & \; $S/N$ & \; z \\[0.5ex]
    & \; [s] & \; [s] & \; [s] & \; [s] & \;  \\[0.5ex] \hline
   \\[1.8ex]
\endfirsthead

%This is the header for the remaining page(s) of the table...
\multicolumn{10}{l}{{\tablename} \thetable{} -- Continued} \\[0.5ex]
  \hline \hline \\[2ex]
   GRB & \; $\Delta t_{\rm min}$ & \; $\Delta t_{\rm snr}$ & \; $T_{\rm 90}$ & \; $T_{\rm R45}$ & \; $\sigma_{X, t_{\rm min}}$ & \; $\sigma_{X, t_{\rm snr}}$ & \; $\chi^2/\nu$ & \; $S/N$ & \; z \\[0.5ex] \hline
  \\[1.8ex]
\endhead

%This is the footer for all pages except the last page of the table...
  \\[1.8ex] \hline
  \multicolumn{10}{l}{{Continued on Next Page\ldots}} \\[0.5ex]
\endfoot

%This is the footer for the last page of the table...
  \\[1.8ex] \hline \hline
\endlastfoot
041220 & \; $1.78\pm0.30$ & \; $0.64\pm0.11$ & \; $3.52\pm0.13$ & \; $1.04\pm0.054$ & \; 0.468 & \; 0.249 & \; 0.0005 & \; 23.5 & \; ... \\[0.5ex]
041223 & \; $0.32\pm0.05$ & \; $0.12\pm0.02$ & \; $108.00\pm0.52$ & \; $19.2\pm0.22$ & \; 0.255 & \; 0.118 & \; 1.119 & \; 177.1 & \; ... \\[0.5ex]
041224 & \; $2.51\pm0.43$ & \; $1.00\pm0.48$ & \; $141.48\pm1.89$ & \; $27.54\pm0.71$ & \; 0.253 & \; 0.134 & \; 1.099 & \; 54.9 & \; ... \\[0.5ex]
041228 & \; $9.90\pm1.68$ & \; $3.54\pm0.60$ & \; $55.50\pm3.39$ & \; $16.28\pm0.442$ & \; 0.302 & \; 0.155 & \; 0.478 & \; 42.2 & \; ... \\[0.5ex]
050117 & \; $ ... $ & \; $0.31\pm0.22$ & \; $168.96\pm0.86$ & \; $47.52\pm0.854$ & \; ... & \; 0.179 & \; 9.758 & \; 78.4 & \; ... \\[0.5ex]
050124 & \; $0.64\pm0.11$ & \; $0.23\pm0.14$ & \; $3.36\pm0.13$ & \; $1.08\pm0.033$ & \; 0.359 & \; 0.190 & \; 2.067 & \; 36.9 & \; ... \\[0.5ex]
050128 & \; $ ... $ & \; $0.18\pm0.09$ & \; $24.48\pm3.63$ & \; $4.48\pm0.187$ & \; ... & \; 0.297 & \; 3.583 & \; 41.6 & \; ... \\[0.5ex]
050215A & \; $ ... $ & \; $5.58\pm2.64$ & \; $66.50\pm0.99$ & \; $6\pm0.429$ & \; ... & \; 0.621 & \; 3.452 & \; 10.2 & \; ... \\
\end{longtable}
\tablecomments{
Burst durations ($T_{\rm 90}$ and $T_{\rm R45}$), $S/N$ values, and redshifts were taken from the \citet{2007ApJ...671..656B} catalog \citep[see, also][]{2010ApJ...711..495B,2013AstRv...8a.103B}. 
Bursts where linear phases in $\sigma_{X,\Delta t}$ could not be fitted by a model with confidence level $>$ 90\%, are considered to yield upper 
limits $\Delta t_{\rm min} \le  \Delta t_{\rm snr}$.}
\end{scriptsize}
\end{center}

\end{document}